\documentclass[aps,showpacs,amsmath,amssymb,reprint,twocolumn,graphicx,superscriptaddress]{revtex4}
\usepackage{graphicx}
\usepackage{dcolumn}
\usepackage{latexsym,bm}
\usepackage{multirow}

\begin{document}
\title{Flat Chern Band in a Two-Dimensional Organometallic Framework}

\author{Zheng Liu}
\affiliation{Department of Materials Science and Engineering, University of Utah, Salt Lake City, UT 84112, USA}

\author{Zheng-Fei Wang}
\affiliation{Department of Materials Science and Engineering, University of Utah, Salt Lake City, UT 84112, USA}

\author{Jia-Wei Mei}
\affiliation{Institute for Theoretical Physics, ETH Zurich, 8093 Zurich, Switzerland}

\author{Yong-Shi Wu}
\affiliation{State Key Laboratory of Surface Physics and Department of Physics, Fudan University, Shanghai 200443, China}
\affiliation{Department of Physics and Astronomy, University of Utah, Salt Lake City, UT 84112, USA}

\author{Feng Liu}
\email{fliu@eng.utah.edu}
\affiliation{Department of Materials Science and Engineering, University of Utah, Salt Lake City, UT 84112, USA}

\date{\today}
\begin{abstract}
By combining exotic band dispersion with nontrivial band topology, an interesting type of band structure, namely the flat chern band (FCB), has recently been proposed to spawn high-temperature fractional quantum hall states. Despite the proposal of several theoretical lattice models, however, it remains a doubt whether such a ``romance of flatland'' could exist in a real material. Here, we present a first-principles design of a two-dimensional (2D) Indium-Phenylene Organometallic Framework (IPOF) that realizes a nearly FCB right around the Fermi level by combining lattice geometry, spin-orbit coupling and ferromagnetism. An effective four-band model is constructed to reproduce the first-principles results. Our design in addition provides a general strategy to synthesize topologically nontrivial materials in virtue of organic chemistry and nanotechnology.

\end{abstract}


\maketitle

Whenever an unconventional band structure is brought out from a conceptual model into a real-world material, a wide range of theoretical advances and technological innovations will be triggered. The well-known examples are graphene \cite{RMP11Geim, RMP11Novoselov} and topological insulators \cite{RMP10Hasan,
RMP11Qi}, which are featured with exotic dispersive bands and nontrivial topological bands, respectively. By combining these two features, another interesting type of band, namely the flat Chern band (FCB) has recently been proposed \cite{PRL11Neupert, PRL11Sun, PRL11Tang}, which is dispersionless and characterized by a nonzero Chern number. A well-defined FCB requires its band width smaller than both the energy gap between the FCB and other bands and the interaction energy scale \cite{PRL11Neupert, PRL11Sun, PRL11Tang}. Since the kinetic energy is strongly quenched, carriers in the FCB experience strong Coulomb interaction in addition to the topological frustration that in together spawn unprecedented topological strongly-correlated electronic states \cite{NatCom11Sheng,PRL11Venderbos,PRL11Wang}. Fundamentally different from the narrow bands commonly existing in heavy fermion compounds, which has a trivial Chern number, the outcome of a FCB requires a delicate balance of lattice hopping, SOC and ferromagnetism. Because of the stringent criteria, no real material to date has been experimentally observed to contain a FCB.

Advances on synthetic chemistry and nanotechnology have shown the potential in producing complex lattices \cite{AngW09Sakamoto,Nat05Barth}. Recent experiments using substrate-mediated self-assembly have successfully fabricated 2D organometallic frameworks with different lattice symmetry. For example, Shi et al. \cite{JACS11Shi} have synthesized two 2D frameworks by co-depositing tripyridyl molecules with Fe and Cu atoms on Au (111) surface, which form Kagome and triangular lattices, respectively. These covalent organic frameworks are found to exhibit remarkable thermal stability. In this Letter, we present a first-principles design to realize the FCB in a 2D Indium-Phenylene Organometallic Framework (IPOF). Density functional theory (DFT) calculations \cite{SM} show that in this unique IPOF structure, a nearly flat band appears around the Fermi level characterized by a nontrivial $Z_2$ topological number. Upon p-type doping, the spin degeneracy of the flat band is spontaneously lifted; the ferromagnetic flat band is then found to have a nontrivial Chern number.

Figure 1a shows the atomic structure of IPOF, whose key feature is to bind p-oribtal heavy elements (In) with organic ligands (para-phenylenes) into a hexagonal lattice. Each hexagonal unit cell contains three phenylenes and two In atoms. As a common feature of group-III elements, the In atoms naturally bond to three phenylenes in a planar triangular geometry. Such a bond configuration is identical to that in triphenyl-indium $In(C_6H_5)_3$, a common indium compound \cite{JAC40Gilman, organicIn85}. A similar organic framework assembled from 1,4-benzeneddiboronic acid on Ag (111) surface has been fabricated in experiment, with only In atoms replaced by boroxine rings \cite{JACS08Zwaneveld,PRB11Ourdjini}. The stability of IPOF has been examined by first-principles lattice relaxation and phonon calculations; details are discussed in Sec. II of the Supplementary Material (SM). The proposed synthesis process of IPOF is described in Sec. III of SM.

\begin{figure}[ht]
\includegraphics[width=0.5\textwidth]{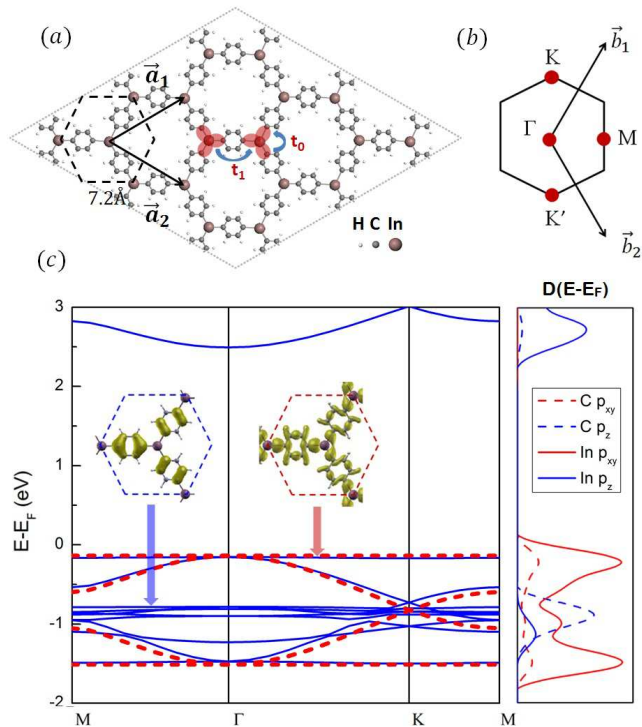}
\caption{ \label{FIG1}(Color online) (a) The atomic structure of IPOF. (b) the first Brillouin zone and special k-points. (c) (Left) Band structure without SOC from DFT (blue solid curves) and model Hamiltonian (eq. (8) with $t_1=0.7eV$,  $\lambda=0$, $M=0$ (red dashed curves). Insets are the wavefunction isosurfaces of two states denoted by arrows. (Right) Atomic orbital projected density of states. }
\end{figure}

To examine the electronic structure of IPOF, we first purposely exclude SOC from the calculation. The resulting electronic band structure, wavefunction and atomic-orbital projected density of states (DOS) are summarized in Fig. 1c. The band structure presents a nonmagnetic insulator picture. Electronic bands at the band edge exclusively come from the p-orbitals of C and In atoms. Specifically, at the valence edge, there are ten bands within $2eV$ from the valence band maximum, which can be divided into two groups. Around $E-E_F=1eV$, there are six $p_z$ bands, which can be attributed to the highest  $\pi$-electron molecular levels from the three benzene rings, as shown by both the wavefunction and the DOS. The remaining four bands exhibit the in-plane $p_{xy}$ features, which arise from hopping among the  $\sigma$-bonds between In and C atoms (red shaded ellipses in Fig. 1a). The top and bottom bands in the four $p_{xy}$ bands are nearly flat in the whole Brillouin zone with a narrow bandwidth of $10meV$ (without SOC). The middle two dispersive bands form a Dirac cone at the $K$ points similar to graphene. The flat bands touch the dispersive bands at the $\Gamma$  point. In contrast to the benzene molecular levels, the wavefunction of the $p_{xy}$ flat bands surprisingly has a distribution across the whole lattice, i.e. they are not localized states (insets of Fig. 1c). This is the first indication that these flat bands are unique and nontrivial.

Next, we include SOC in the calculation, and the results are shown in Fig. 2a. Comparing Fig.2a with Fig. 1c, the most significant difference is that several degenerate points of the $p_{xy}$ bands, e.g. the $\Gamma$ and $K$ points, are split. Consequently, the two flat bands become separated from the dispersive bands. The direct ($\Delta_{dir}^{12}$) and indirect gap ($\Delta_{ind}^{12}$) between the top flat band and the nearest dispersive band (Fig. 2c) are  $\Delta_{dir}^{12}=90meV$ and  $\Delta_{ind}^{12}=30meV$. The separation between the top and bottom flat bands ( $\Delta^{14}$ in Fig. 2a) is $1.4eV$. Another effect of the SOC splitting is that the band width ($W$) of the flat bands increases from $10$ to $60meV$. Note that all the bands are still spin degenerate because of the time-reversal and inversion symmetry. The topology of the flat band is characterized by a nontrivial Z2 toplogical number as discussed in Sec. IV of SM.

Partially-filled flat band is unstable due to the large DOS at the Fermi level. It has been rigorously proved that even an arbitrarily small Coulomb interaction will drive the system into a ferromagnetic ground state at specific filling factors \cite{JPA91Mielke, JPA91Mielke2}. To obtain some insights within the DFT formalism, we dope the system by manually reducing the number of valence electrons in the unit cell, while maintaining the charge neutrality with a compensating homogeneous background charge. This makes the top flat band partially filled. Calculations reveal a spontaneous spin polarization perpendicular to the 2D plane. Figs. 2b and 2c shows the band structure of this ferromagnetic ground state under doping. The spin-up and spin-down bands are separated apart, with the Fermi level shifting below the topmost spin-polarized flat band. This ferromagnetic ground state is $5meV$ lower than the spin-unpolarized state in total energy, which can be further stabilized by applying an external Zeeman field. The spin splitting, $U\sim100meV$ (Fig. 2b), represents the strength of the on-site Coulomb interaction. We thus estimate the electron-electron interaction of the $p_{xy}$ electrons in IPOF to be on the same order. For comparison, the spin splitting of the $p_z$ electrons (benzene molecular levels) is one order of magnitude smaller.

\begin{figure}[ht]
\includegraphics[width=0.5\textwidth]{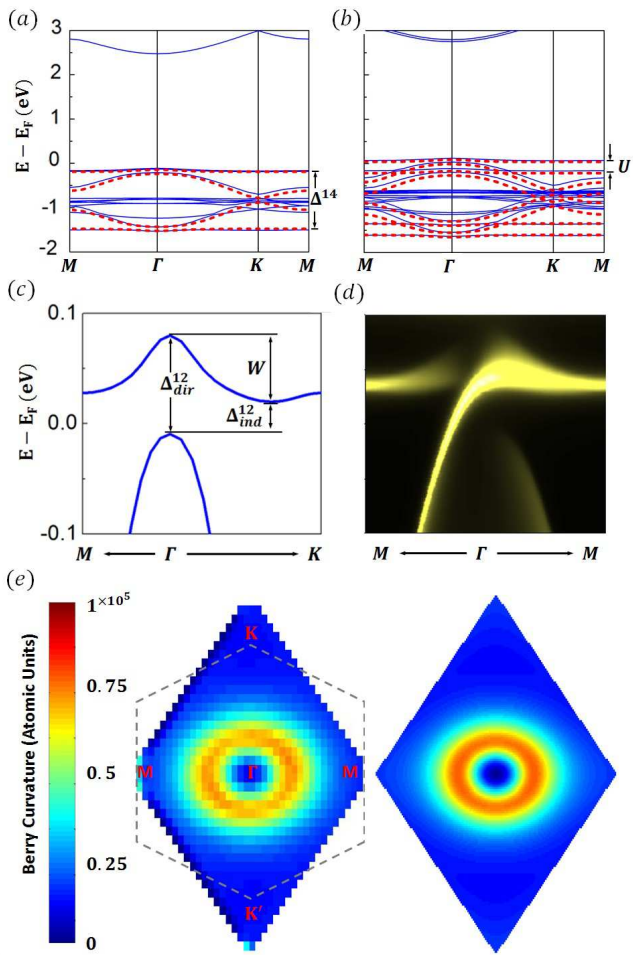}
\caption{\label{FIG2}(Color online)(a) Band structure with SOC. (b) Band structure with SOC when doping one hole into the unit cell. The solid (blue) curves are DFT results. The dashed (red) curves are from model Hamiltonian eq. (8) with parameters (a) $t_1=0.7eV$, $\lambda=0.05eV$, $M=0$; (b) $t_1=0.7 eV$, $\lambda=0.05 eV$, $M=0.1eV$  (c) Zoomed-in band plot around the Fermi level. (d) Momentum-resolved edge density of states of a semi-infinite IPOF with SOC and doping. The brightness is proportional to the magnitude of the density of states.  (e) Distribution of the Berry curvature $F_12(\textbf{k})$ of the top flat band in the reciprocal space. Left is the DFT result; right is the model result using the same parameters as in (b). }
\end{figure}

We will now examine the topology of the topmost spin-polarized flat band by directly calculating its Chern number based on its DFT wavefunctions. The Chern number is defined as:

\begin{equation}
c=\frac{1}{2\pi i}\int_{BZ} d^2\textbf{k} F_{12}(\textbf{k}),
\end{equation}

where the Berry curvature $F_{12}(\textbf{k})$ is given by:

\begin{eqnarray}
F_{12}&=&\partial_1 A_2(\textbf{k})-\partial_2 A_1(\textbf{k})\nonumber\\
A_i&=&\langle u_\textbf{k}|\partial _i|u_\textbf{k}\rangle
\end{eqnarray}

$|u_\textbf{k}\rangle$ is the Bloch function of the flat band and the derivative $\partial_i$ stands for $\partial / \partial_{k_i}$. The distribution of Berry curvature is shown in Fig. 2e, exhibiting an interesting ring pattern. The integration of the Berry curvature in the whole Brillouin zone gives $c=1$. Therefore, we conclude that this top flat band is a FCB.

A direct manifestation of the Chern number is the number of chiral edge modes circulating around the boundary. To check the edge property, we have calculated the momentum-resolved edge DOS of a semi-infinite IPOF \cite{SM}. As shown in Fig. 2d, one single edge band emerges in the gap ($\Delta^{12}$) and unidirectionally connects the top flat band with the dispersive band below it, signifying the typical feature of a chiral mode and indicating the Chern number $c=1$.

We have further constructed an effective Hamiltonian to understand the origin of the FCB in IPOF. Considering that the flat band comes from hopping among the $\sigma$  bonds, we use these  $\sigma$-bond orbitals as basis to construct the effective Hamiltonian. There are three  $\sigma$ bonds around each In atoms as marked by the red elliptical shades in Fig. 1a. We first symmetrize the basis orbitals by considering the nearest-neighbor (NN) hopping ($t_0$) among the three bonds:

\begin{eqnarray}
H_0=t_0\left(
      \begin{array}{ccc}
        0 & 1 & 1 \\
        1 & 0 & 1 \\
        1 & 1 & 0 \\
      \end{array}
    \right)
\end{eqnarray}

The eigenstates of $H_0$ consist of a doublet ($\epsilon_d$) and a singlet ($\epsilon_s$):
\begin{eqnarray}
\epsilon_d&=&t_0 \qquad \{\begin{array}{c}
                   \Psi_{d1}=\frac{1}{\sqrt{2}}[1,0,-1]^T \\
                   \Psi_{d2}=\frac{1}{\sqrt{6}}[1,-2,1]^T
                 \end{array}\nonumber\\
\epsilon_s&=&-2t_0 \qquad \Psi_s=\frac{1}{\sqrt{3}}[1,1,1]^T
\end{eqnarray}

The symmetrized eigenstates transform like vectors: $\Psi_{d1}$  and $\Psi_{d2}$ within the $x$-$y$ plane; $\Psi_s$ perpendicular to the $x$-$y$ plane. As long as $\Psi_s$ is far away from the doublet, it can be reasonably neglected. The problem is then reduced to a four-dimensional subspace spanned by two sets of $\Psi_{d1}$  and $\Psi_{d2}$ associated with the two In atoms in the unit cell. Considering the next-NN (NNN) hopping ($t_1$) via the benzene rings, we can write out a four-band hopping Hamiltonian for each spin:

\begin{eqnarray}
  H_{hop}=-t_1\left(
                \begin{array}{cccc}
                  0 & 0 & V_{xx} & V_{xy} \\
                  0 & 0 & V_{xy} & V_{yy} \\
                  V_{xx}^* & V_{xy}^* & 0 & 0 \\
                  V_{xy}^* & V_{yy}^* & 0 & 0 \\
                \end{array}
              \right), 
\end{eqnarray}
in which $V_{xx}=\frac{1}{2}(1+e^{i\textbf{k}\cdot \textbf{a}_1})$ , $V_{xy}=\frac{\sqrt{3}}{6}(1-e^{i\textbf{k}\cdot \textbf{a}_1})$  and $V_{yy}=\frac{1}{\sqrt{6}}(1+e^{i\textbf{k}\cdot \textbf{a}_1}+4e^{i\textbf{k}\cdot \textbf{a}_2})$  ;  $\textbf{a}_{1,2}$ is the lattice vector as shown in Fig. 1a. Without loss of generality, we have set the on-site energy of $\Psi_{d1}$  and $\Psi_{d2}$ as zero. The eigenvalues of $H_{hop}$ give four energy bands (the red dashed curves in Fig. 1c), which agree well with the DFT results. $H_{hop}$ effectively describes a honeycomb lattice with two in-plane vector-like orbitals at each site, i.e. the $p_{xy}$-orbital counterpart of graphene. As discussed in Refs. \cite{PRL07Wu, PRB08Wu}, this kind of hopping Hamiltonian guarantees two completely flat bands, because their Wannier functions forbid any electron ``leakage'' to outside through a destructive interference.

With regard to SOC, we start from the original atomic form: $H_{SOC}=\lambda \hat{L}\cdot\hat{\sigma}$. By realizing that $\Psi_{d1, d2, s}$ transforms like three vectors, we can write out its second quantization form:
\begin{eqnarray}
H_{soc}=\lambda(ic_{d1\downarrow}^\dagger c_{d2\downarrow}-ic_{d1\uparrow}^\dagger c_{d2\uparrow}+c_{s\uparrow}^\dagger c_{d1\downarrow}-c_{s\downarrow}^\dagger c_{d1\uparrow}\nonumber\\
+ic_{s\uparrow}^\dagger c_{d2\downarrow}-ic_{s\downarrow}^\dagger c_{d2\uparrow}+h.c.),
\end{eqnarray}
in which $c_{i\sigma_z}^\dagger$  creates an electron with spin $\sigma_z$ on the $\Psi_i$ orbital.
To reduce  $H_{SOC}$ into the four band ($\Psi_{d1,d2}$) subspace, we keep only the leading terms:
\begin{eqnarray}
H_{soc}^0=\lambda(ic_{d1\downarrow}^\dagger c_{d2\downarrow}-ic_{d1\uparrow}^\dagger c_{d2\uparrow}),
\end{eqnarray}
which do not include coupling between different spin components, so the spin-up and spin-down spaces get automatically separated. For each spin subspace, the SOC can also be written as a $4\times4$ matrix:
\begin{eqnarray}
H_{soc,\sigma_z}^0=\sigma_z\lambda\left(
                                    \begin{array}{cccc}
                                      0 & -i & 0 & 0 \\
                                      i & 0 & 0 & 0 \\
                                      0 & 0 & 0 & -i \\
                                      0 & 0 & i & 0 \\
                                    \end{array}
                                  \right)
,
\end{eqnarray}
where $\sigma_z=\pm 1$ is the spin eigenvalue. This SOC represents an imaginary on-site coupling, independent of momentum. It pins an additional phase to the electrons when they hop between the two on-site orbitals and result in nontrivial topology when electrons travel in the lattice via $H_{hop}$.

The spontaneous magnetization under doping is a many-body effect \cite{JPA91Mielke, JPA91Mielke2}. Within the single-electron picture, we reproduce the ferromagnetism by simply adding a Zeeman term. Finally, the effective Hamiltonian to describe the single-electron property of IPOF can be written as:
\begin{eqnarray}
H_{eff,\sigma_z=\pm 1}(k)=H_{hop}(k)+H_{soc,\sigma_z}^0+H_{Zeeman} \nonumber \\
=-t_1\left(
                \begin{array}{cccc}
                  0 & 0 & V_{xx} & V_{xy} \\
                  0 & 0 & V_{xy} & V_{yy} \\
                  V_{xx}^* & V_{xy}^* & 0 & 0 \\
                  V_{xy}^* & V_{yy}^* & 0 & 0 \\
                \end{array}
              \right)
              + \sigma_z\lambda\left(
                                    \begin{array}{cccc}
                                      0 & -i & 0 & 0 \\
                                      i & 0 & 0 & 0 \\
                                      0 & 0 & 0 & -i \\
                                      0 & 0 & i & 0 \\
                                    \end{array}
                                  \right) \nonumber \\
              + \sigma_z M
\end{eqnarray}

There are three parameters in the model: the NNN hopping parameter $t_1$, the spin-orbit coupling strength $\lambda$ , and the spontaneous magnetization $M$. Note that the NN hopping term $t_0$ does not explicitly enter the four-band model that renders the flat band more accessible in IPOF than in other model lattices \cite{PRL11Neupert, PRL11Sun, PRL11Tang}. The higher-order hopping terms, e.g. the next-NNN hopping will affect the flatness, but they are already negligible in IPOF, smaller than 10 meV according to the DFT results. By fitting the three parameters to the DFT results, the dispersion of the four $p_{xy}$ bands can be well reproduced, as shown in Fig. 1c, 2a and 2b. To ensure that this simple model also reproduces the wavefunction property, we in addition calculate the Berry curvature using the model eigenstates; the result is also in good agreement with the DFT result (Fig. 2e). The middle two dispersive bands from the model are found to have $c = 0$; the bottom flat band has $c =  1$; thus the total Chern number of the four band subspace is zero. These justifications confirm the validity of the effective Hamiltonian, which will be useful for future investigations going beyond the DFT formalism.

In summary, we reiterate in Tab. I several key energy scales associated with the FCB in IPOF. Further studies including many-body effects are necessary to determine the ground state and low-energy excitation of the carriers in the FCB. The emergence of fractional quantum hall state requires further enlarging the energy gap $\Delta^{12}$ and reducing the bandwidth $W$ to satisfy the condition  $\Delta\gg U>W$ \cite{PRL11Neupert,PRL11Tang, NatCom11Sheng}. To achieve this intriguing state, the organic nature of IPOF provides the flexibility to tune the parameters in kinds of chemical ways, e.g. by functionalizing the benzenes with different chemical groups, replacing the benzenes with other organic ligands, or using different metal atoms. Furthermore, similar design principles can also be applied to realize other exotic band structures. For example, by simply replacing In with Bi and Mn, one makes new organic topological insulator and magnetic topological insulator, respectively \cite{submit12Wang}.

\begin{table}[ht]
\caption{Energy scales associated with the FCB in IPOF}
\begin{ruledtabular}
\begin{tabular}{cccc}
  Property & Symbol & Value & Ref. \\
  \hline
  Band width & $W$ & $60meV$ & Fig.2c \\
  Spin splitting & $U$ & $100meV$ & Fig.2b \\
  Debye temp. & $\omega_d$ & $300meV$ & Fig.S1 \\
  Energy gap & $\Delta^{12}_{dir}$ & $90meV$  & Fig.2c \\
   & $\Delta^{12}_{ind}$ & $30meV$ & Fig.2c \\
  & $\Delta^{14}$ & $1.4eV$ & Fig.2a \\
\end{tabular}
\end{ruledtabular}
\end{table}

ZL, ZFW and FL acknowledge support from the DOE-BES (DE-FG02-03ER46027). ZFW also acknowledges support from ARL. JWM acknowledges support from the Swiss National Funds. YSW is supported in part by U.S. NSF Grant No. PHY-0756958.

\end{document}